\documentclass[aps,prl,twocolumn,floatfix]{revtex4}
\usepackage{bm}
\usepackage{epsf}
\usepackage{amssymb}
\usepackage{amsmath}
\usepackage{graphicx}
\usepackage{rotating}
\usepackage{epsfig}
\usepackage{psfrag}
\usepackage{amsmath}
\usepackage{hyperref}
\newcommand{\bk}{{\bf k}}
\newcommand{\x}{{\rm x}}

\begin{document}

\title{Molecular Pairing and Fully-Gapped Superconductivity in Yb doped CeCoIn$_5$}

\author{Onur Erten$^1$, Rebecca Flint$^2$, Piers Coleman$^{1,3}$}
\affiliation{$^1$Center for Materials Theory, Rutgers University, Piscataway, New Jersey, 08854, USA \\ $^2$Department of Physics and Astronomy, Iowa State University, 12 Physics Hall, Ames, Iowa 50011 USA \\ $^3$Department of Physics, Royal Holloway, University of London, Egham, Surrey TW20 0EX, UK }

\begin{abstract}
The recent observation of fully-gapped superconductivity in Yb doped
CeCoIn$_5$ poses a paradox, for the disappearance of nodes suggests that they
are accidental, yet d-wave symmetry with protected nodes is we ll established
by experiment.  Here, we show that composite pairing provides a natural
resolution: in this scenario, Yb doping drives a Lifshitz transition of the
nodal Fermi surface, forming a fully-gapped
d-wave molecular superfluid of composite pairs.  The $T^4$ dependence of the
penetration depth associated with the sound mode of this condensate
is in accordance with observation.
\end{abstract}
\maketitle

{\sl Introduction:} CeCoIn$_5$ is an 
archetypal 
heavy fermion
superconductor with $T_c = 2.3 K$\cite{Petrovic_JPCM2001}. The
Curie-Weiss susceptibility signaling unquenched local moments,
persists down to the superconducting transition
\cite{Petrovic_JPCM2001}. Local moments, usually harmful to
superconductivity actually participate in the condensate and a
significant fraction of the local moment entropy ($0.2-0.3 \log2$ per
spin) is quenched below $T_c$.

The behavior of this material upon Yb doping is quite unusual: 
the depression of superconductivity with doping is extremely
mild with an unusual  linear dependence of the 
transition temperature $T_c({\rm x})=T_c(0)\times(1-\x)$, where $\x$ is the Yb
doping\cite{Shu_PRL2011}. Moreover, recent
measurements\cite{Tanatar_2014} of  the temperature dependent 
London penetration depth $\Delta\lambda(T)$ suggest that the nodal d-wave
superconductivity (where $\Delta\lambda(T) \sim T-T^2$) becomes fully
gapped ($\Delta\lambda(T) \sim T^n, n\gtrsim3$) beyond a critical Yb
doping $\x_{c}\sim0.2$. Normally the disappearance of nodes would suggest
that they are accidental,  as in $s^{\pm}$ superconductors. However
directional probes of the gap, 
including scanning tunneling spectroscopy
(STM)\cite{Davis_NatPhys2013, Yazdani_NatPhys2013}, thermal conductivity measurements in
a rotating
magnetic field\cite{Izawa_PRL2001} and torque magnetometry\cite{Maple_PRB2008} 
strongly  suggest that pure CeCoIn$_{5}$ is 
a d-wave superconductor with symmetry-protected nodes. 
{\sl How then, can a nodal d-wave superconductor become
fully-gapped upon doping?}

\begin{figure}[t!]
\centerline{
\includegraphics[width=8.5cm]{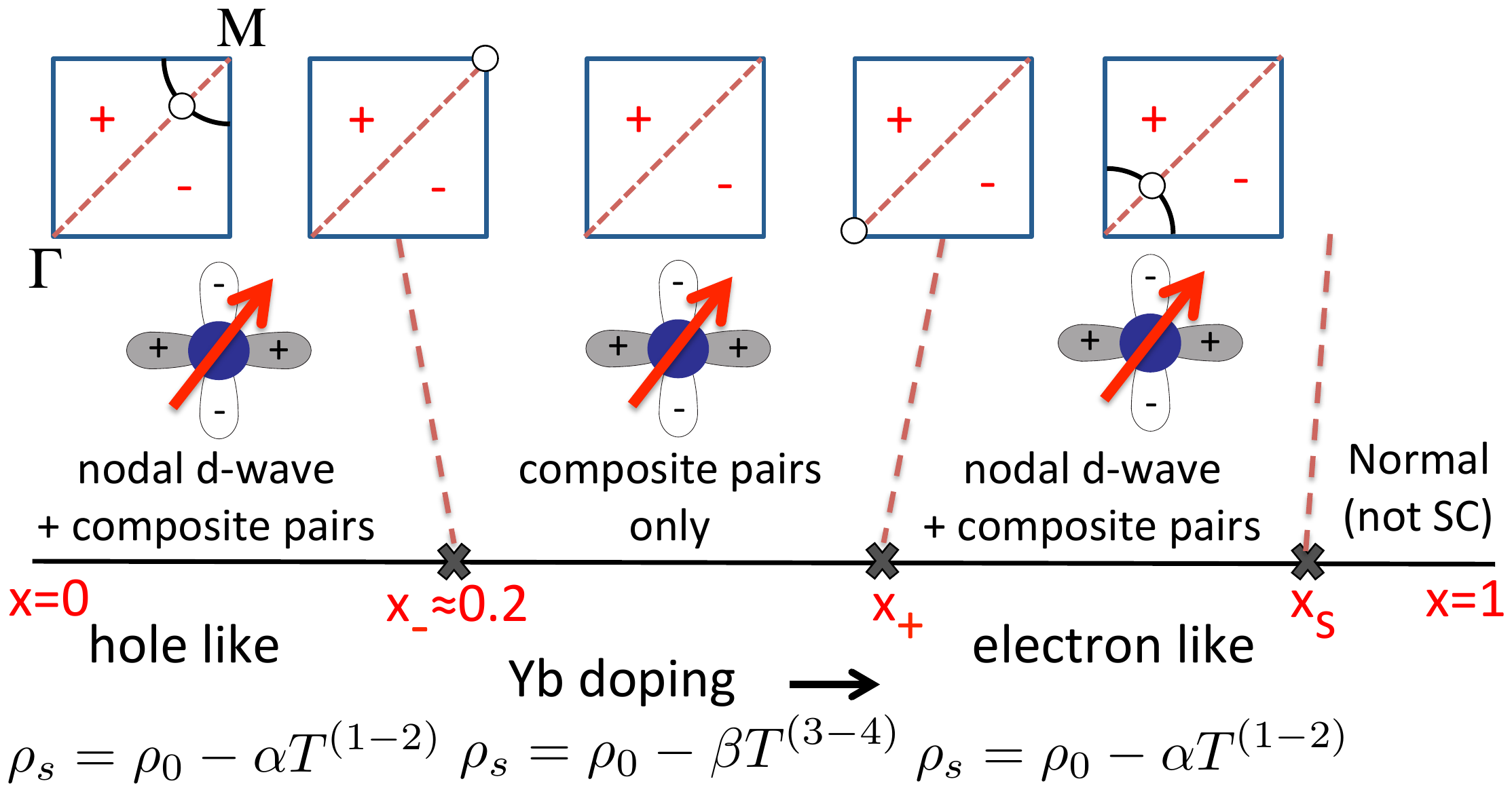}
}
\caption{Schematic phase diagram for the Yb$_{x}$Ce$_{1-x}$CoIn$_5$. For $x<0.2$ the temperature dependence of London penetration depth $\Delta\lambda \sim T-T^2$, consistent with nodal d-wave superconductivity in clean and dirty limits respectively. However for $x>0.2$, the power law of the $\Delta\lambda$ exceeds 2 and reaches up to 4\cite{Tanatar_2014}. This is incompatible with nodal d-wave superconductivity and suggests a fully-gapped state. We argue that in the gapless phase Cooper pairs and composite pairs coexist whereas in the fully gapped phase, only the composite pairs are present. As function of Yb doping, chemical potential increases and the nodes of the order parameter moves to the corners of the Brillouin zone and annihilate. We predict that upon further doping, there is a second quantum phase transition to a reentrant gapless phase. Upon even further doping, there might be a third transition to a normal state at $x_s$ since superconductivity has been only observed up to $x\leq 0.65$.}
\label{Fig:1}
\end{figure}

Here we provide a possible resolution of this paradox, 
presenting a mechanism by which nodal superconductors can become
fully gapped systems without change of symmetry, through the formation
of composite pairs.
A composite d-wave superconductor contains two 
components: a d-wave BCS condensate and a molecular 
superfluid of d-wave composite pairs\cite{Coleman_PRB1999}. Here we show when
the scattering phase shift off the magnetic ions is tuned via doping,
a Lifshitz transition occurs which removes the nodal heavy Fermi
surface, without losing the superfluid stiffness, revealing an underlying molecular superfluid of
d-wave composite pairs (see Fig.~\ref{Fig:1}).

In the absence of an underlying Fermi surface, 
a composite paired superconductor 
can be regarded as
Bose-Einstein condensate of weakly interacting, 
charge 2e d-wave bosons in
which the Bogoliubov quasiparticle spectrum is fully
gapped\cite{Randeria_PRB1990}, with a residual 
linear sound mode with dispersion $E_q\sim 
v_{s}q$, 
cut off by the plasma frequency $\omega_{p}\sim v_{s}/\lambda_{L}$
at wavevectors 
below the inverse penetration depth $q<< 1/\lambda_{L}$.
At temperatures above 
the plasma frequency, the superfluid stiffness 
is governed by Landau's two-fluid theory of superfluids, in which the
excitation of the normal superfluid 
is predicted to give rise to a power law dependence of the penetration
depth $\Delta \lambda(T) \sim T^4$ in three dimensions, consistent
with experiments\cite{Tanatar_2014}.

A quantum critical point recently observed for  $\x\sim \x_{c}$ in
transverse magnetoresistance measurements \cite{Hu_PNAS2013} appears to
coincide with the disappearance of the superconducting nodes. 
At larger Yb doping, we expect a second quantum critical point into 
a reentrant 
gapless phase as shown in Fig.~\ref{Fig:1} with the 
redevelopment of a d-wave paired 
heavy  electron pocket around the $\Gamma$ point 
in the Yb rich Kondo lattice. 

We now expand on the idea of composite pairing and discuss
its detailed application to Yb doped CeCoIn$_5$ and the
consequences and the predictions of our theory.
\begin{figure}[t!]
\centerline{
\includegraphics[width=7cm]{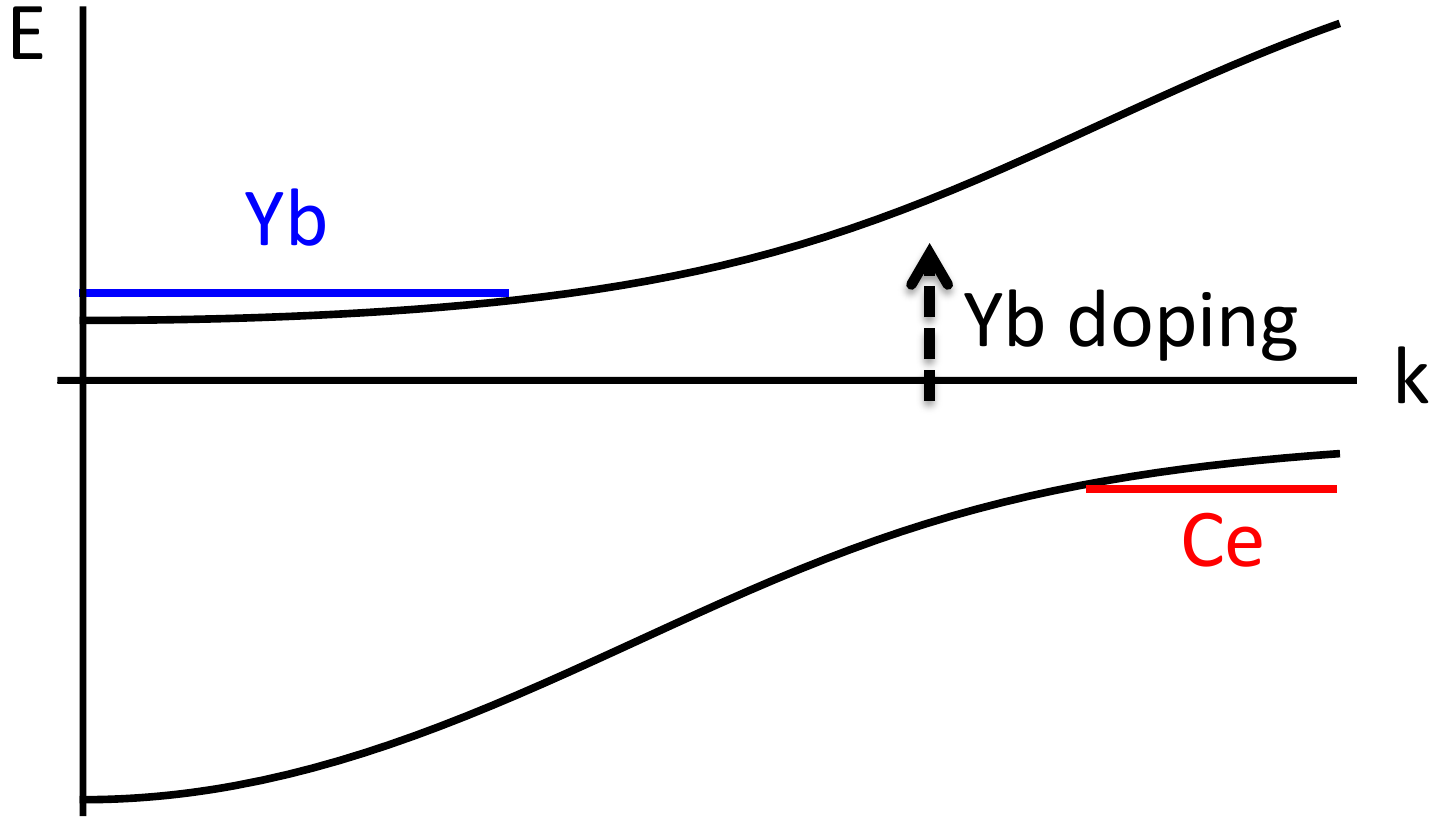}
}
\caption{Schematic heavy fermion band structure. Ce Kondo lattice is hole-like whereas Yb Kondo lattice is electron-like. Thus upon Yb doping, chemical potential increases and heavy fermion band structure turns from hole-like to electron-like.
}
\label{Fig:2}
\end{figure}

{\sl Composite pairing:} The 
composite pairing concept was first
introduced in the context of odd-frequency pairing
\cite{Abrahams_PRB1995}, and later associated with 
the composite binding of a  Cooper pairs 
with local moments\cite{Coleman_PRB1999, Flint_NatPhys2008,
Flint_PRL2010, Flint_PRB2011}. Various other forms of composite pairing
have been recently suggested in the context of cuprate superconductors\cite{Berg_PRL2010}.
Composite pairing naturally emerges
within a two-channel Kondo lattice model where constructive 
interference between two 
spin-screening channels drives to local pairing. 
Composite pairing can be alternatively regarded
as an intra-atomic version of the 
resonating valence bond pairing mechanism\cite{Anderson_Science1987, Kotliar_PRB1988}.
The composite pair amplitude is given by
\begin{eqnarray}
\Lambda_C(j) = \langle \psi^\dagger_{1j} \vec{\sigma}(i\sigma_2)\psi^\dagger_{2j} \cdot \vec{S}_{f} (j) \rangle
\end{eqnarray}
where $\psi^\dagger_{\Gamma j}$ creates conduction electrons in the
Wannier state of channel $\Gamma \in (1,2)$ and $\vec{S}_{f} (j)$ describes 
the spin operator of the local f-moment at site $j$.  So, here two conduction electrons in orthogonal channels are screening the same local moment, giving rise to a singlet composite pair, which exists within a single-unit cell and thus can be regarded as a molecular unit.  The
$\psi$'s can be decomposed into plane waves using the relation 
$\psi_{\Gamma j
\sigma}=\sum_{\bk } \Phi^{\Gamma}_{\sigma \sigma^\prime} (\bk ) c_{\bk 
\sigma^\prime} e^{i {\bf k}\cdot {\bf R}_j}$ where the form factor
$\Phi^{\Gamma}_{\sigma \sigma^\prime} (\bk )$ captures the different symmetries of the two types of hybridization.  While in a simple model, one can take $\Phi_{1k}$ and $\Phi_{2k}$ to be s-wave and d-wave, in real materials the momentum dependence will be more complicated, and the $\Phi_{\Gamma k}$'s become matrices that are only diagonal in the
absence of spin-orbit coupling.  
The Kondo coupling in the two $\Gamma$ channels $J_{\Gamma}$ 
are a consequence of virtual charge fluctuations 
from the singly occupied ground state into the excited
empty and doubly occupied states. 

The symmetry of the composite pair condensate is determined by
the product of the two form factors $\Phi_{1k}\Phi_{2k}$. In the simple
model, where $\Phi_{1k}$ and $\Phi_{2k}$ have s-
and d-wave symmetries respectively, the composite pairs will have a d-wave symmetry. A more detailed analysis involving the underlying crystal field symmetries finds that the two channels have $\Gamma_6$
and $\Gamma_7$ symmetries, again leading to $d_{x^2-y^2}$-wave like composite pairs\cite{Flint_PRB2011}. 
The
superfluid stiffness
\begin{equation}\label{}
Q=Q^{BCS}+Q^{M}
\end{equation}
has two components\cite{Coleman_PRB1999}: a BCS component
\begin{equation}\label{}
Q^{BCS} = \frac{n_s e^2}{m^*}
\end{equation}
derived from the paired heavy electron fluid, where $n_s$ is the superfluid density, and a composite component
\begin{eqnarray}
Q^M \simeq
\sum_k \frac{\Lambda_C^2(\Phi_{1\bk}\nabla \Phi_{2\bk}- \Phi_{2\bk}\nabla \Phi_{1\bk})^2} {\Sigma_N^2\sqrt{\epsilon_\bk^2+2\Sigma_N^2}}\sim \frac{e^2}{\hbar^2 a} (k_B T_c),
\end{eqnarray}
here given at zero temperature, resulting from the mobility of the molecular pairs and
derived ultimately from the non-local character (momentum dependence)
of the hybridization form factors.  
Here, $a$ is the lattice constant,
$\Sigma_N$ is proportional to the normal (hybridization) part of the
conduction electron self-energy and $\epsilon_\bk$ is the conduction
electron dispersion. 
$Q_{M}$ is directly proportional to the
condensation energy, a consequence of ``local pair''
condensation and it does not depend on the presence of a Fermi surface.
In three dimensions, $Q^{BCS} \sim (\epsilon_F/a)(e/\hbar)^2$ is proportional to the Fermi
energy: in conventional superconductors the superfluid stiffness is much
greater than $T_c$ and the BCS component will dominate, but 
as the Fermi surface shrinks, $Q^{BCS}$ vanishes. Normally, this would
drive a superconductor-insulator transition,
but now the superconductivity is sustained
by the additional stiffness $Q_{M}$ of the composite pair condensate. Note that, within this picture, the BCS and composite components have the same origin and do not compete with one another.  

For example, consider a single channel Kondo lattice model at half
filling, for which the ground state is a Kondo insulator with a gap to
quasiparticle excitations. 
The inclusion of a  second Kondo channel leads to
composite pairing beyond a critical ratio of the coupling
constants. (There is no Cooper instability in this case since
there is no Fermi surface.) As a result the Bogoliubov quasiparticle
spectrum is fully-gapped even though the composite order parameter has
d-wave symmetry. This state is an example of a Bose-Einstein
condensate of d-wave molecules.

{\sl Connection with Yb doped CeCoIn$_5$:}  Due to the 
tetragonal crystal field, the low lying physics of CeCoIn$_{5}$ 
is governed by a
low lying $\Gamma_7$ Kramers doublet\cite{Willers_PRB2010}. 
The Kondo effect in Ce and Yb heavy fermion compounds results from 
high frequency valence fluctuations. 
In Ce compounds 
the dominant valence fluctuations occur 
between the $4f^{1}$ and $4f^{0}$ configuration 
$4f^{1}\leftrightharpoons
4f^{0}+e^{-}$, giving rise to an average f-occupation below unity 
($n_f^{Ce} \sim 0.9$)\cite{Booth_PRB2011,Dudy_PRB2013}. 
Using the Friedel sum rule, this 
gives rise to a scattering phase shift
$\delta  < \frac{\pi}{2}$ and in the lattice, to 
hole-like heavy Fermi surfaces. 
By contrast, Yb heavy fermion materials involve 
valence fluctuations between the $4f^{13}$ and $4f^{14}$ configurations
$e^{-}+4f^{13}\leftrightharpoons 4f^{14}$, so the average 
f-occupation of the active Kramers doublet exceeds one
($n_f^{Yb} \sim 1.7$)\cite{Booth_PRB2011,Dudy_PRB2013}
\footnote{Note that there is a rapid change in the valence for $x<0.2$
which we ignore for simplicity.}, the corresponding  scattering 
phase shift $\delta > \frac{\pi}{2}$ and an electron-like Fermi
surface in the Kondo lattice (Fig.~\ref{Fig:2}).
As the Yb doping proceeds, 
the typical character of the resonant scattering 
changes from Cerium-like to
Ytterbium-like and the occupancy of the low-lying magnetic doublet 
$n_f$ will increase as a function of Yb doping
\begin{eqnarray}
n_f(\x) &\approx& (1-\x)n_f^{Ce} +\x n_f^{Yb}\\
&&0.9+0.8\x
\end{eqnarray}
Yb doping effectively
increases the f-electron count $n_{f}$, 
causing the average scattering phase shift $\delta$
to rise. As a function of doping,
the nodes of the gap move to the zone corner as shown in
Fig. 1 (a), and annihilate once 
$\delta \sim  \pi/2$, forming
a Kondo insulator immersed within a composite d-wave superfluid.

STM quasiparticle interference experiments show that there are 
two hole-like bands and an electron-like band\cite{Yazdani_NatPhys2013}, where the superconducting gap has been identified on the hole-like bands. We 
predict that the annihilation of these 
nodes as a function of doping will be 
seen on these bands in particular in the $\alpha$ band as defined in ref. \cite{Davis_NatPhys2013}.  Indeed
the disappearance an electron-like band is seen both in ARPES\cite{Dudy_PRB2013} and dHvA\cite{Polyakov_PRB2012}
experiments. However, the behavior of the electron-like band is unclear.
At higher doping, the nodes should reappear. 
(see Fig.~\ref{Fig:1}). Indications of strong
Fermi surface reconstructions seen in transport data 
around $\x=0.55$~\cite{Polyakov_PRB2012}  may be tentatively 
identified with this second quantum critical point. 
It would be interesting to see if the nodal quasiparticles reappear beyond this point.

\begin{figure}[t!]
\vspace{0.2cm}
\centerline{
\includegraphics[width=7.5cm]{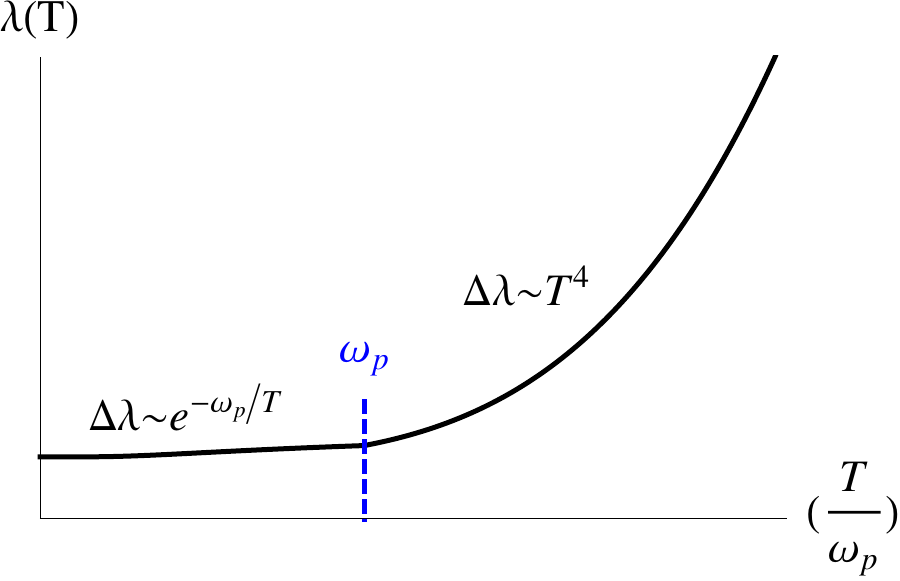}
}
\vspace{.2cm}
\caption{Temperature dependence of the London penetration depth. For $T<\omega_p$, the $\Delta\lambda(T)$ is exponentially suppressed, whereas it crosses over to $T^4$ for $T>\omega_p$.}
\label{Fig:3}
\end{figure}

{\sl Penetration depth:} In nodal superconductors, the temperature
dependence of the change of the penetration depth $\Delta
\lambda(T)\sim T^n$ is either $n=1$ in the clean limit or $n=2$ in the
dirty limit. A higher power is inconsistent with a nodal gap.
Experiments\cite{Tanatar_2014} show that $n\sim 3-4$ for $\x\sim
0.2$. By contrast, the
temperature dependent penetration depth of a fully-gapped
molecular condensate is governed by the superfluid 
sound mode whose scale is set
by the superfluid stiffness $Q_C$. 
In a Landau two fluid model, the temperature dependence of the superfluid
density is 
\begin{eqnarray}
\rho_s(T)&=&\rho_0-\frac{(2e)^2}{d}\int \frac{d^dq}{(2\pi)^d}\Big(-\frac{\partial n(\omega_q)}{\partial \epsilon_q} \Big)\left(\frac{q}{m^{*}} \right)^{2}
\end{eqnarray}
where $d$ is the dimension and $m^{*}$ is the effective mass of the
composite pairs. Since $[q]=T$, by power counting $\rho_{s}\propto
T^{d+1}$, leading to 
a temperature dependence of the penetration depth given by 
$\lambda(T)=\lambda_0+\beta T^n$ where $n=d+1$. 
Since the condensate is charged, 
the linear sound spectrum will be gapped by the plasma frequency
$\omega_p$. We estimate $\omega_p$ to be about 10-100 mK \footnote{The
plasma frequency is set by ratio of the speed
of light in the medium to the penetration depth $\omega_p =
c^*/\lambda$. The mobility of the composite pairs results from 
the momentum dependence of the form factors and the scale is set by the hybridization $c^*\sim \sqrt{T_KW}/a$ 
where $T_K$, $W$, $a$ are the Kondo temperature, conduction electron bandwidth and the lattice constant respectively. 
We used $T_K \sim 5$ K, $W \sim 10^4$ K, $a \sim 0.2$ nm and $ \lambda \sim 300 $ nm, which leads to $\omega_p \sim 100$ mK.} 
and assuming the composite superfluid is three dimensional, the
temperature dependence of the penetration depth will have a power law
$n=4$ for $T>\omega_p$ as shown in Fig~\ref{Fig:3}, which is
consistent with experiments.  We should note that this power law is
not uniquely identified with composite pairing, as the Gorter-Casimir
two-fluid behavior of s-wave superconductors also gives a $T^4$
dependence at low temperatures.  However, composite pairing 
provides an explanation for the transition from nodal d-wave to
nodeless superconductivity within a single pairing mechanism.

{\sl Resistivity above $T_c$:} An over-simplistic application of Tinkham's fluctuation conductivity theory~\cite{Tinkham_1996} to our case, gives a resistivity of the form $\rho(T)=\rho(T_c)+A T^{(4-d)/2}$ giving a $T^{1/2}$ power-law in three dimensions which reflects the phase space for superconducting fluctuations. Remarkably, experiments~\cite{Hu_PNAS2013}  display a robust $T^{1/2}$ resistivity at dopings $x>0.2$, surviving over a decade in temperature up to 20K. However, this temperature range is far too great to be attribute to fluctuations
about a weak-coupling BCS superconductor. One possibility is that the vicinity to unitary pairing enhances the range of the superconducting fluctuations. Alternatively, critical two-channel Kondo impurity physics  may play a role in reinforcing the robust
$T^{1/2}$ resistivity, in accordance with composite pairing.



{\sl Thermal conductivity:} The thermal conductivity, $\kappa$, of a
d-wave superconductor is dominated by the nodal quasiparticle
excitations which leads to a linear temperature
dependence\cite{Durst_PRB2000} with a coefficient that oscillates
in a perpendicular magnetic field as a function of in-plane
orientation\cite{Izawa_PRL2001}. In the fully-gapped phase,
the linear temperature dependence of $\kappa$ will be exponentially suppressed $\kappa/T \sim (\Delta/T)^2 e^{-\Delta/T}$, leading
to a jump in $\kappa_0/T$ at the Lifschitz transition and $\kappa \sim T^3$ phonon behavior.  The oscillations of
$\kappa$ in magnetic field will also be suppressed due to the absence
of nodal quasiparticles, and it should show activated behavior.



{\sl Conclusion:} Composite pairing provides a natural explanation for
the development of a 
fully-gapped state in Yb doped CeCoIn$_5$.  As a function of Yb
doping the chemical potential increases and the nodes move to the
corner of the Brillouin zone.  When the phase shift reaches $\pi/2$,
the nodes annihilate, completely depleting the Fermi
surface. The resulting 
fully-gapped state has a superfluid stiffness derived from the
composite pairs, a form of ``molecular'' condensate. 
The predicted sound mode as the low energy excitation of the
molecular condensate may be observable in ultrasound experiments.
Moreover, the
unusual linear doping dependence of the transition temperature 
$T_c$ can now simply understood as the BEC
temperature of the composite pairs.



We also note 
that the mechanism presented here may apply to a 
much broader class of strongly interacting electron fluids: the recent
observation of fully gapped superconductivity 
developing in CeCu$_2$Si$_2$ in a field
\cite{Kittaka_PRL2014} and Ce doped
PrPt$_4$Ge$_{12}$\cite{Huang_arxiv2014} are interesting additional
candidate examples of this phenomenon that deserve future examination. 

{\sl Acknowledgments:} We gratefully acknowledge stimulating conversations with Makariy Tanatar, Ruslan Prozorov, Carmen Almasan, Maxim Dzero,
Gabriel Kotliar and Emil Yuzbashyan. Onur Erten is partly supported by National Science Foundation grant NSF DMR 1308141, the David and Lucile Packard Foundation grant 200131769. Piers Coleman is supported by National Science Foundation grant, NSF DMR 1309929.

\end{document}